\documentclass[aps,prd,preprint,superscriptaddress,tightenlines,nofootinbib,showpacs]{revtex4}

\usepackage{graphicx}
\usepackage{dcolumn}
\usepackage{bm}

\begin{document}

\preprint{CLNS 08/2041}       
\preprint{CLEO 08-23}         
\def\etaP{\eta^{\prime}}

\title{\boldmath New Measurement of
Exclusive Decays of the $\chi_{c0}$ and $\chi_{c2}$ to Two-Meson Final States }

\author{D.~M.~Asner}
\author{K.~W.~Edwards}
\author{J.~Reed}
\author{A.~N.~Robichaud}
\author{G.~Tatishvili}
\affiliation{Carleton University, Ottawa, Ontario, Canada K1S 5B6}
\author{R.~A.~Briere}
\author{H.~Vogel}
\affiliation{Carnegie Mellon University, Pittsburgh, Pennsylvania 15213, USA}
\author{P.~U.~E.~Onyisi}
\author{J.~L.~Rosner}
\affiliation{Enrico Fermi Institute, University of
Chicago, Chicago, Illinois 60637, USA}
\author{J.~P.~Alexander}
\author{D.~G.~Cassel}
\author{J.~E.~Duboscq}
\thanks{Deceased}
\author{R.~Ehrlich}
\author{L.~Fields}
\author{R.~S.~Galik}
\author{L.~Gibbons}
\author{R.~Gray}
\author{S.~W.~Gray}
\author{D.~L.~Hartill}
\author{B.~K.~Heltsley}
\author{D.~Hertz}
\author{J.~M.~Hunt}
\author{J.~Kandaswamy}
\author{D.~L.~Kreinick}
\author{V.~E.~Kuznetsov}
\author{J.~Ledoux}
\author{H.~Mahlke-Kr\"uger}
\author{D.~Mohapatra}
\author{J.~R.~Patterson}
\author{D.~Peterson}
\author{D.~Riley}
\author{A.~Ryd}
\author{A.~J.~Sadoff}
\author{X.~Shi}
\author{S.~Stroiney}
\author{W.~M.~Sun}
\author{T.~Wilksen}
\affiliation{Cornell University, Ithaca, New York 14853, USA}
% ...
\collaboration{CLEO Collaboration}
\noaffiliation

\date{November 3, 2008}

\begin{abstract}
Using a sample of $2.59 \times 10^7$ $\psi(2S)$ decays collected
by the CLEO--c detector, we present
results of a study of $\chi_{c0}$ and $\chi_{c2}$  decays
into two-meson final states.
We present the world's most precise measurements 
of the 
$\chi_{cJ,(J=0,2)} \to \pi^+\pi^-$,
$\pi^0\pi^0$, $K^+K^-$, 
$K^0_SK^0_S$, $\eta\eta$
and
$\eta^{\prime}\eta^{\prime}$
branching fractions,
and a search for $\chi_c$ decays into $\eta\eta^{\prime}$.
These results shed light on the mechanism of charmonium decays
into pseudoscalar mesons.

\end{abstract}

\maketitle


The $\chi_{cJ}$ mesons ($J=0,1,2$)
form a triplet of $c\bar{c}$ states with one unit of orbital angular momentum.
They are not produced directly
in $e^+e^-$ annihilations, but the large branching fractions
of $\psi(2S)\to\chi_{cJ}\gamma$ make $e^+e^-$ collisions
at the $\psi(2S)$ energy a very clean  
environment for $\chi_{cJ}$ investigation. 

The $\chi_{cJ}$ mesons decay into a wide variety 
of different
multi-hadron
states. Of these, the two-meson states have the benefit of being
comparatively straightforward to detect and to model theoretically.
However, theoretical models based on the color singlet model make
predictions well below the data, even when the parameters of the 
model are stretched to extremes \cite{Brambilla}.
Recent theoretical work has focussed on 
the Color Octet Model (COM) \cite{COM}, 
whereby contributions from the sub-process
$c\bar{c}g \to q \bar{q}$ 
are
taken into account. This source has been 
shown to be a possible mechanism to make up the deficit \cite{Kroll}.
However, these theoretical results were made with a very rough model for 
the color-octet contribution to the wave-function of the $\chi_c$.
Fits \cite{Zhao} to the existing 
measurements \cite{PDG} indicate a reasonable theoretical 
understanding of the processes involved, but data remain sparse and
theoretical uncertainties are still 
large. 
The study of the
decays to higher mass mesons ($\eta$ and $\etaP$), 
offers the possibility of investigating the
contribution of doubly-OZI 
suppressed decays (DOZI), which may compete with singly-OZI
suppressed decays (SOZI) \cite{Zhao},
and may also contribute to the understanding of the structure of 
the $\eta$ and $\etaP$ mesons \cite{Thomas}.
The $\chi_{c1}$ cannot decay into two pseudoscalar mesons because 
of spin-parity conservation, so we do not consider it
further. 

The data were taken by the CLEO-c detector \cite{CLEOC} 
operating at the Cornell 
Electron Storage Ring (CESR) with $e^+e^-$ collisions at 
a center of mass energy corresponding to the $\psi(2S)$
mass of 3.686 GeV/$c^2$. The data corresponds
to an integrated luminosity of  
56.3 ${\rm pb^{-1}}$ and
the total number of $\psi(2S)$ events, 
determined according to the
method described in \cite{PSI2S},
is calculated as $(2.59\pm 0.05) \times 10^7$.

Photons were detected using the 
CsI crystal calorimeter \cite{CLEOII}, which has an energy resolution of
2.2\% at 1 GeV, and 5\% at 100 MeV. Photon candidates were required to have 
a lateral shower shape consistent with that expected for a photon, and
to be not matched with any charged track.
We combine two photon candidates to make $\pi^0$ candidates. These are then kinematically
constrained to the $\pi^0$ mass and those combinations within 3 standard deviation
of this mass are retained for further analysis.

Charged particles were detected in a drift chamber system immersed in 1.0 T 
solenoidal magnetic field. The solid angle for detecting charged particles was
93\% of 4$\pi$, and the resolution 0.6\% at 1 GeV.
To discriminate charged kaons from charged pions, we combined specific ionizations
$(dE/dx)$ measured in the drift chamber and log-likelihoods obtained
from the ring-imaging \v Cerenkov detector (RICH) \cite{RICH} to form a log-likelihood difference: 
${\cal L}(K-\pi)={\cal L}_{\mathrm RICH}(K)-{\cal L}_{\mathrm RICH}(\pi)+
\sigma^2_{dE/dx}(K)-\sigma^2_{dE/dx}(\pi)$, where negative ${\cal L}(K-\pi)$
implies the particle is more likely to be kaon than a pion. For all charged
kaons we require ${\cal L}(K-\pi)<0$ and ${\cal L}(K-p)<0$,
and for charged pions we require
${\cal L}(\pi-K)<0$ and ${\cal L}(\pi-p)<0$.
As there is potential contamination from lepton pairs in 
the $\pi^+\pi^-$ final state, we use the muon chamber and CsI information
and
require that at least one of the tracks in this mode is
inconsistent with being due to either a muon or an electron

We use three decay modes for $\eta$ detection, $\gamma\gamma$, $\pi^+\pi^-\pi^0$,
and $\pi^+\pi^-\gamma$, and two modes for $\etaP$ detection, $\eta\pi^+\pi^-$ and 
$\gamma\pi^+\pi^-$. In each case we combine the four-momenta of the decay 
products into an $\eta^{(\prime)}$ candidate, kinematically constrain the candidate to 
its nominal mass and retain those candidates with a fit $\chi^2 < 9/1$ degree of freedom.

For events with two distinct meson candidates, 
we combine the candidates into a 
$\chi_{c}$ candidate. At this stage of the analysis, 
the invariant mass resolution of the $\chi_c$ is 
approximately 
15 MeV$/c^2$.
We then search for any unused photon in the event 
and add that to the $\chi_c$ candidate to form
a $\psi(2S)$ candidate. This $\psi(2S)$ candidate is then kinematically 
constrained to the four-momentum of the beam, 
the energy of which is calculated using the known
$\psi(2S)$ mass. The momentum is non-zero due to the
finite crossing angle ($\approx 3$ mrad per beam) in CESR. 
To make our final selection, 
we require the $\psi(2S)$ candidate to have a  
$\chi^2$ of less than 25 for the four degrees of freedom for this fit; 
this requirement rejects most background combinations. 
This kinematic fit greatly improves the mass resolution of the 
$\chi_c$ candidate to values ranging from 3.2 to 8.7 MeV$/c^2$
depending on the spin of the $\chi_c$ and the decay mode.

To study the efficiency and resolutions, 
we generated Monte Carlo samples
for each $\chi_c$ into each final state 
using a GEANT-based detector simulation \cite{GEANT}.
The simulated events have an angular distribution of
$(1+\alpha \cos^2 \theta)$, where $\theta$ is the 
radiated photon angle relative to the
positron beam direction, and $\alpha=$ 1  and 1/13 for the 
$\chi_{c0}$, and $\chi_{c2}$ respectively, 
in accordance with expectations for an E1 transition.
The decay products of the $\chi_{c0}$ are generated with a flat
angular distribution. The products of the $\chi_{c2}$ are generated
according to a double correlation function of the polar angles of the 
mesons measured in the $\chi_c$ rest frame relative to the
transition photon direction \cite{DC}.
The efficiencies are shown in Table~I. The efficiencies for 
$\eta$ and $\etaP$ modes include the relevant branching fractions.
\begin{table}[htb]
\caption{
Yields found in the data sample and detection efficiencies
obtained from analyses of Monte Carlo generated events.
}

\begin{tabular}{l|cc|cc}
\hline
\hline
Mode  & \multicolumn{2}{c|}{ $\chi_{c0}$ } & 
\multicolumn{2}{c}{ $\chi_{c2}$} \\
\hline
      & Yield & Efficiency(\%) &  Yield & Efficiency(\%) \\
\hline
$\pi^+\pi^-$ & $8934\pm 111$ & $58.7\pm 2.4$ & $2543\pm56$ & $66.2\pm 2.7$    \\
$\pi^0\pi^0$ & $2807\pm62$ & $40.0\pm 4.4$ & $793\pm 33$ & $48.5\pm 5.3$  \\
$K^+K^-$ & $8156\pm100$ & $53.8\pm 2.5$ & $1645\pm42$ & $60.2 \pm 2.8$    \\
$K^0_SK^0_S$ & $2109\pm49$ & $25.3\pm 1.1$ & $373\pm20$ & $29.3 \pm 1.3$    \\
$\eta\eta$   & $930\pm 35$ & $12.3\pm 1.1$ & $156\pm 14$ & $12.6\pm 1.1$   \\
$\eta\etaP$  & $35\pm13$    & $9.2\pm 0.8$  & $3.3\pm8.0$ & $10.5\pm 0.9$  \\
$\etaP\etaP$  & $413\pm24$    & $8.2\pm 0.6$  & $12\pm7$ & $8.8\pm 0.5$  \\
\hline
\hline
\end{tabular}
\end{table}
                                                                                    
The final invariant mass distributions are shown in Fig.~1.
These distributions are then fit
with two signal shapes comprising Breit-Wigner
functions convolved
with Gaussian resolutions, together with a constant background term. 
The masses and widths
of the Breit-Wigner functions 
were fixed according to the Particle Data Group (PDG) averages \cite{PDG}, and 
the widths of the Gaussian resolution functions were 
fixed at the values found from Monte Carlo simulation. 
The yields from these binned likelihood fits are tabulated in Table~I.

To convert the yields to branching fractions, we divide by the 
product of the number of $\psi(2S)$ events in the data sample,
the detector efficiency,
and the 
branching fractions for $\psi(2S)$ into $\chi_{cJ}$. For the 
last factor 
we use the CLEO measurements of $\mathcal{B}(\psi(2S)\to\gamma\chi_{c0})=
(9.22\pm0.11\pm0.46$)\% and
$\mathcal{B}(\psi(2S)\to\gamma\chi_{c2})=
(9.33\pm0.14\pm0.61$)\% 
\cite{CLEOresult}. The results are tabulated in 
Table~II.

We consider systematic uncertainties from 
many different sources.
All modes have a 2\% uncertainty from the total number of $\psi(2S)$ decays \cite{PSI2S}.
The requirement on the $\chi^2$ of the
constraint to the beam four-momentum has been checked by changing 
the cut value in the range 12--50 and noting
the change in the yield in these, and other similar decay modes.
Based on
this study we place a systematic uncertainty of 2.5\% on the efficiency 
of this 
requirement. The uncertainties due to track reconstruction are 0.3\% per charged track
(0.67\% for kaons). 
The limited Monte Carlo statistics introduces an uncertainty which is in all cases 
less than 1.5\%.
The systematic uncertainty due 
to the photon detection and shower-shape criteria 
is set at 2\% per photon both for the transition photon and for the 
decay products of $\eta$ and $\pi^0$ decays. 
In the cases including $\eta$ decays, this 
contribution is incorporated taking into account the fraction of those
decays that proceed through each $\eta$ decay mode.
The final signal plots are all well fit
using the functions described above.
By studying the variation of the yields of the high statistics modes
resulting from floating the signal parameters, we assign a 2\%
uncertainty in each mode due to the uncertainties in the fitting procedure.
We have checked that the yields from the various decay modes of the $\eta^{(\prime )}$
mesons are consistent with their branching fractions and efficiencies.
When calculating the final branching
fractions, we add the above systematic uncertainties in quadrature.
The uncertainty due to the
$\psi(2S)\to\gamma\chi_c$ branching fractions is kept separate and quoted
as a second systematic uncertainty.

For evaluating the limits in the cases where there is no
significant signal, we take the probability density function and convolve this 
with Gaussian systematic uncertainties. We then find the  
branching fraction that includes 90\% of the total area.
\begin{table}[htb]
\caption{Branching fraction results (in units of $10^{-3}$) for each decay mode.
The uncertainties are statistical, systematic due to this measurement, and systematic
due to the $\psi(2S)\to\chi_{cJ}\gamma$ rate, respectively. 
The limits on the branching fractions
include all systematic uncertainties, and central values for those measurements 
are included in parentheses.}
\begin{tabular}{lccc}
\hline
\hline
Mode  &  & $\chi_{c0}$ &  $\chi_{c2}$ \\
\hline
$\pi^+\pi^-$ 
 & $\ $ This Work$\ $ & $\ \ 6.37\pm0.08\pm0.29\pm0.32\ \ $ &  $\ \ 1.59\pm0.04\pm0.07\pm0.10\ \ $ \\ 
 & PDG \cite{PDG}     & $4.87\pm0.40$               &  $1.42\pm0.16$  \\

$\pi^0\pi^0$       
 & $\ $This Work$\ $ & $2.94\pm0.07\pm0.32\pm0.15$ & $0.68\pm0.03\pm0.07\pm0.04$ \\  
 & PDG    & $2.43\pm .20$             &  $0.71\pm0.08$  \\

$K^+K^-$ 
  &$\ $ This Work$\ $ & $6.47\pm0.08\pm0.33\pm0.32$ & $1.13\pm0.03\pm0.06\pm0.07$ \\
  & PDG      &  $5.5\pm0.6$   & $0.78\pm0.14$  \\

$K^0_SK^0_S$ 
 &$\ $ This Work$\ $ & $3.49\pm0.08\pm0.17\pm0.17$ & $0.53\pm0.03\pm0.03\pm0.03$  \\
 & PDG & $2.77\pm0.34$ & $0.68\pm0.11$ \\

$\eta\eta$
 &$\ $ This Work$\ $ & $3.18\pm0.13\pm0.31\pm0.16$ & $0.51\pm0.05\pm0.05\pm0.03$ \\
 & PDG        &    $2.4\pm0.4$ & $<0.5$  \\

$\eta\etaP$
 &$\ $ This Work$\ $  & $<0.25$  & $<0.06$ \\
 &                    & ($0.16\pm0.06\pm0.01\pm0.01$)& ($0.013\pm0.031\pm0.001\pm0.001$)\\
&PDG & $<0.5$&$<0.26$ \\
 
$\etaP\etaP$ &
 $\ $ This Work$\ $ & $2.12\pm0.13\pm0.18 \pm0.11$ & 
$<0.10$ \\
& & & $(0.056\pm0.032\pm0.005\pm0.003$)\\
& PDG & $1.7\pm0.4$& $<0.4$\\
\hline
\hline
\end{tabular}
\end{table}

Our results are summarized in Table~II, and compared with the PDG
fits \cite{PDG} to results from BES \cite{BES} and CLEO \cite{CLEOD}. These fits
explicitly assume that ${B}(\chi_c\to \pi^+\pi^-)=2{B}(\chi_c\to\pi^0\pi^0)$.
Our results do not include that constraint, but the data are consistent with this
isospin symmetry. Our results are also consistent with the expected result that 
${B}(\chi_c\to K^+K^-)=2{B}(\chi_c\to K^0_SK^0_S)$, whereas previous 
results had indicated that this might not be so in the $J=2$ case.
The largest deviation from previous results ($\approx 3 \sigma$) is in the case
of $\chi_{c0} \to \pi^+\pi^-$.
In the case of the $\chi_{c2}$,
our limit for the branching fraction into $\eta\etaP$
is below the fit value obtained from previous data by
Qiang Zhao \cite{Zhao}, suggesting that the DOZI decays of the $\chi_{c2}$ may contribute
less than indicated by that phenomenological analysis. We note that there is 
an overlap of datasets in the results presented here and those of our previous
analysis of $\eta^{(\prime)}\eta^{(\prime)}$ decays, and so our new results
should replace rather than augment our previous measurements.

In summary, 
we measure branching fractions for $\chi_{c0}$ and $\chi_{c2}$ 
decays into 
$\pi^0\pi^0$, $\pi^+\pi^-$, $K^+ K^-$,
$K^0_SK^0_S$, $\eta\eta$,
and $\etaP\etaP$.
The decay $\chi_{c2}\to \eta\eta$ is observed for the first time and
in all other cases these measurements are more precise than any previously made. 
These results may be used to test the 
role of the Color Octet Mechanism model of $\chi_{cJ}$ decays. The improved limits 
on decays into $\eta\etaP$ are further proof of the small contributions
made by DOZI decays in this system.

We gratefully acknowledge the effort of the CESR staff
in providing us with excellent luminosity and running conditions.
D.~Cronin-Hennessy and A.~Ryd thank the A.P.~Sloan Foundation.
This work was supported by the National Science Foundation,
the U.S. Department of Energy,
the Natural Sciences and Engineering Research Council of Canada, and
the U.K. Science and Technology Facilities Council.

\begin{figure}[htb]

\includegraphics*[width=5.0in]{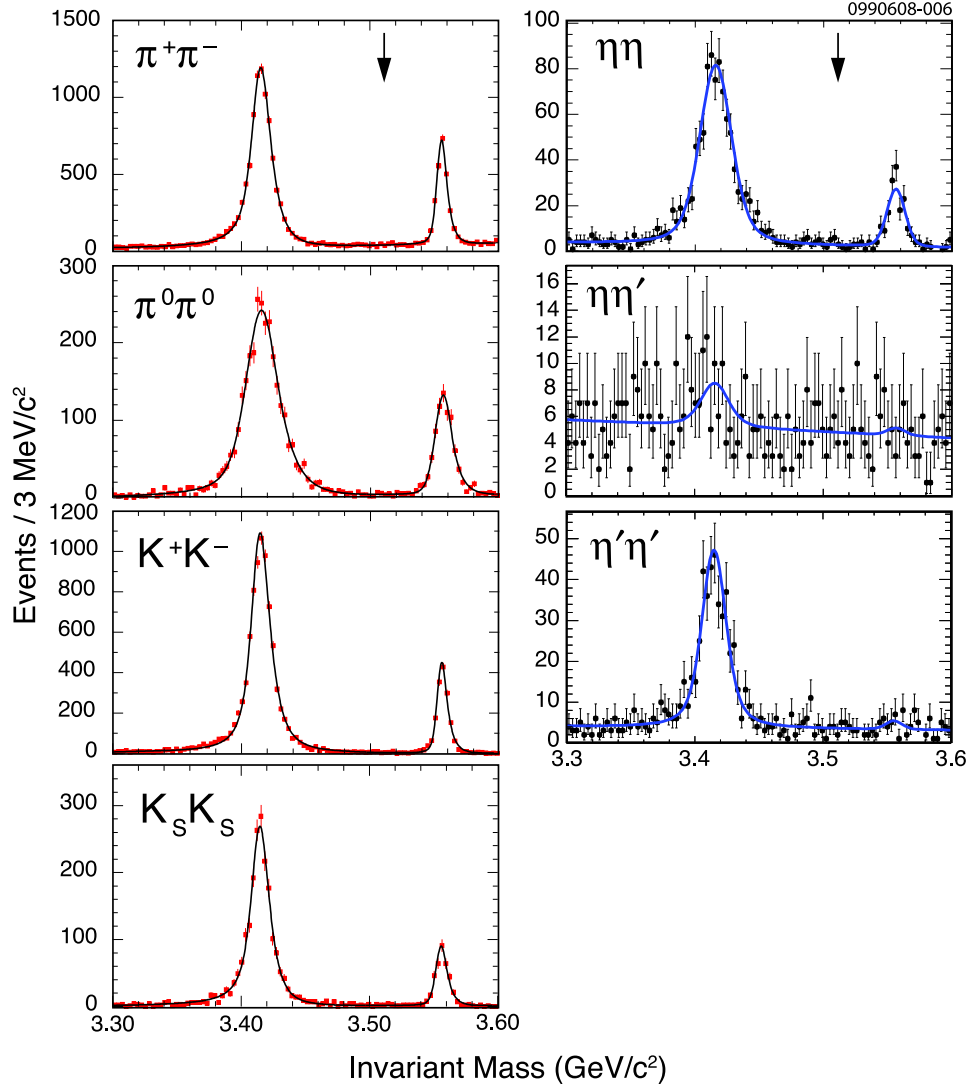}

\caption{Invariant mass distributions for $\pi^+\pi^-$, $\pi^0\pi^0$,
$K^+K^-$, $K^0_SK^0_S$, $\eta\eta$,
$\eta\etaP$, $\etaP\etaP$.  
The fits are described in the text.
The downward arrows are at the value of the invariant mass of the $\chi_{c1}$.
}
\end{figure}

\end{document}